\documentclass[article]{cjaa}    
\usepackage{graphicx}                   
\usepackage{url}

\begin{document}

   \title{Searching for Radio Pulsars in 3EG Sources at Urumqi Observatory}

\volnopage{Vol.6 (2006) , Suppl. 2, 294-297}      

   \setcounter{page}{1}                      

   \author{Jiang Dong\inst{1,2} \mailto{}
           \and  Na Wang\inst{1}
           }
           
   \offprints{Dong Jiang}                    

   \institute{Urumqi Observatory, NAOs, CAS,
           40-5 South BeiJiang Road, Urumqi, 830011, China
      \email{djcosmic@gmail.com}
      \and   Graduate University of Chinese Academy of Sciences}


   \date{Received~~2005 month day; accepted~~2005~~month day}
   \abstract
  {A pulsar searching system has been operating at 18-cm band at the Urumqi Observatory 25-m radio telescope since mid-2005. Test observations for known pulsars show the system can perform pulsar searching. Perspect of using this system to observe    
3EG sources and other target searching are prepared and discussed. 
   }

  \authorrunning{Jiang Dong}            
   \titlerunning{Finding Radio Pulsar in 3EG Sources at Urumqi Observatory}  

   \maketitle


%
%

\section{Introduction}           

Among the more than 1700 known pulsars, seven are seen at $\gamma$-ray 
frequency (Lorimer \& Kramer, 2005). It is worth mentioning that six of 
the seven have detected radio emission (McLaughlin 2001). This might 
indicate the possible relationship between $\gamma$-ray sources and 
radio pulsars (McLaughlin 2001, Lorimer 2003, Gonthier et al. 2002, 
Cheng et al. 2004, Qiao et al. 2004). Encouraged by such phenomenon, 
we started the program of searching for radio pulsar in the error 
boxes of 3EG sources.

The Compton Gamma Ray Observatory (CGRO) was the second of NASA's Great
Observatories which was operated from April 5, 1991 to June 4, 2000. 
The Energetic Gamma Ray Experiment Telescope (EGRET) 
provides the highest energy gamma-ray window for the Compton Observatory. 
Its energy range is from 20~MeV to 30~GeV. EGRET is 10 to 20 times larger
and more sensitive than previous detectors operating at these high energies 
and has made detailed observations of high-energy processes associated 
with diffuse gamma-ray emission, gamma-ray bursts, cosmic rays, 
pulsars, and active galaxies known as blazars. 

In next section we will introduce the pulsar searching system at Nanshan, Urumqi Observatory, and discuss using of small radio telescolpe (SRT) 
to do pulsar search and possible transient radio source search 
(Cordes, J.M., et al., 2004). We report the prime results
and discuss the perspect of future observations in Section 3.

\section{The facilities at Urumqi Observatory}    

The 25-m radio telescope at Nanshan is operated by Urumqi Observatory. It locates close to the geographic center of Asia, with an altitude of 2029\ m above sea level, longitude $87\degr$ and latitude $+43\degr$. 

A pulsar timing system at 18~cm was built in 1999 (Wang et al., 2001). For this band, the telescope has cassegrain focus and uses a horn feed receiving orthogonal linear polarisations. The receiver has dual-polarisation, cryogenic pre-amplifiers with center radio frequency (RF) of 1540~MHz and total bandwidth of 320~MHz. The receiver noise temperature is less than 10~K, and the system temperature is approximately 23~K. The polarisations are amplified and then down-converted to intermediate frequency (IF) in the range 80—400~MHz using a local oscillator (LO) at 1300 MHz. After conversion, the signals are fed to a filterbank system 
which has 128 2.5-MHz channels for each polarisation. The online program of Pulsar Searching Data Acquistion is written in Visual C++ and run in windows NT system. Our sampling interval is  256~$\mu$s or higher so that it is sensitive for detecting millisecond pulsar (MSP) (Roberts et al. 2004). 

After sum two polarisation data in software, we will perform 
a standard Fourier analysis 
use the {\bf Sigproc} 
\footnote{\url{http://sigproc.sourceforge.net/}} (Lorimer,D.,2000) software package
for slow ($> 4$~ms) pulsars 
and a fast-folding algorithm {\bf FFA} 
\footnote{\url{http://www.mpifr-bonn.mpg.de/staff/peter/ffades.en.html}}
for very slow (3-20s) pulsars,
and full acceleration searches 
using the {\bf  Presto}\footnote{\url{http://www.cv.nrao.edu/~sransom/}} search software (Ransom,S.M., 2001) 
which is sensitive to pulsars in tight binary systems. 

\begin{center}
\begin{figure}[htbp]
\includegraphics[width=14cm,height=7cm]{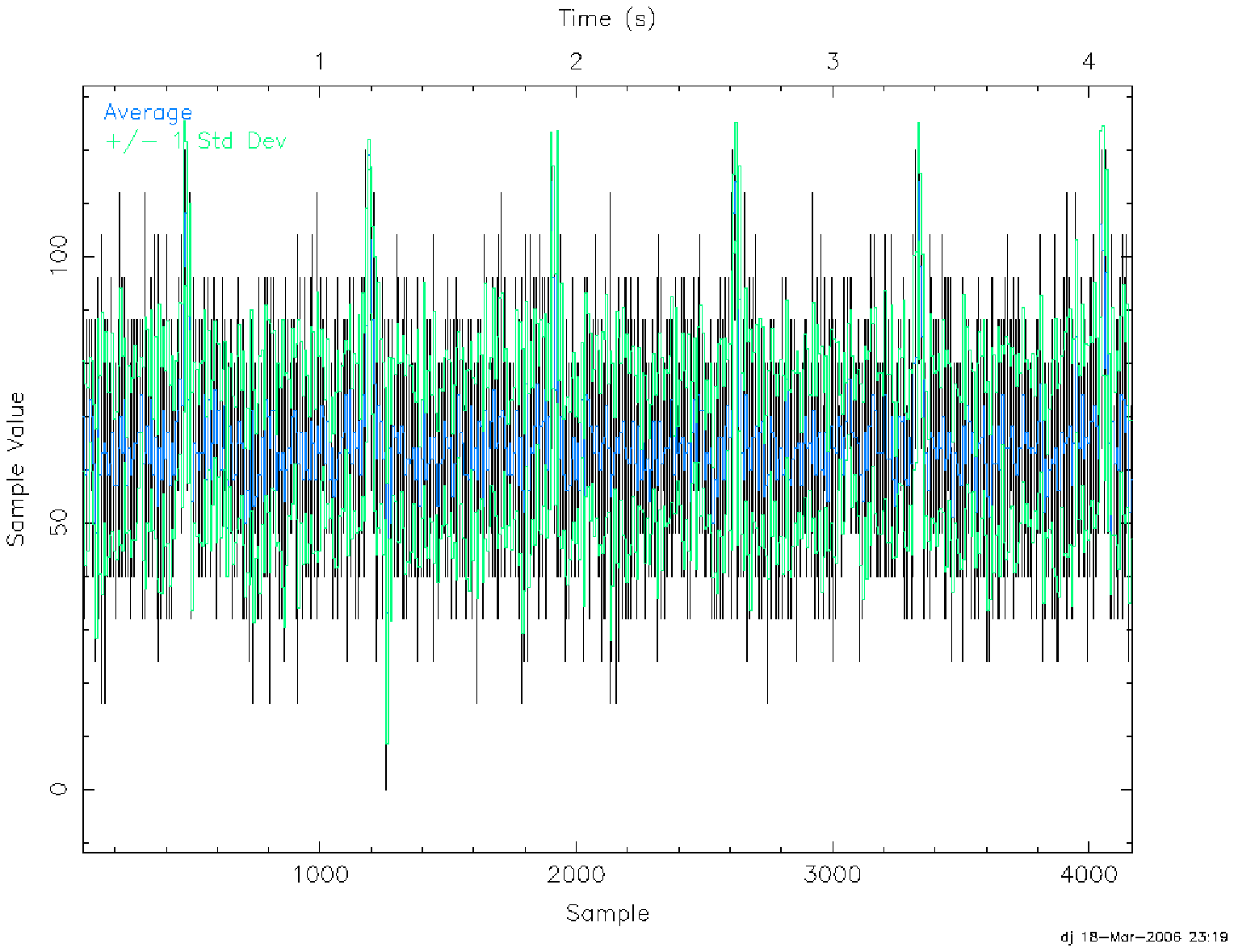}
\\
\caption{De-dispersed data for PSR B0329+54 after summing two polarisations off line.}
\label{t}
\end{figure}
\end{center}

The technique of resample has been applied in pulsar search (Lorimer \& Kramer, 2005). When the ratio of oversampling equals to 64 for 1-bit quantiser, we almost can achive the precision of 14 bits in the decimator (Oppenheim, Schafer \& Buck, 1999). We will use it in our search. 

\begin{center}
\begin{figure}[htbp]
\includegraphics[width=12cm, height=7cm]{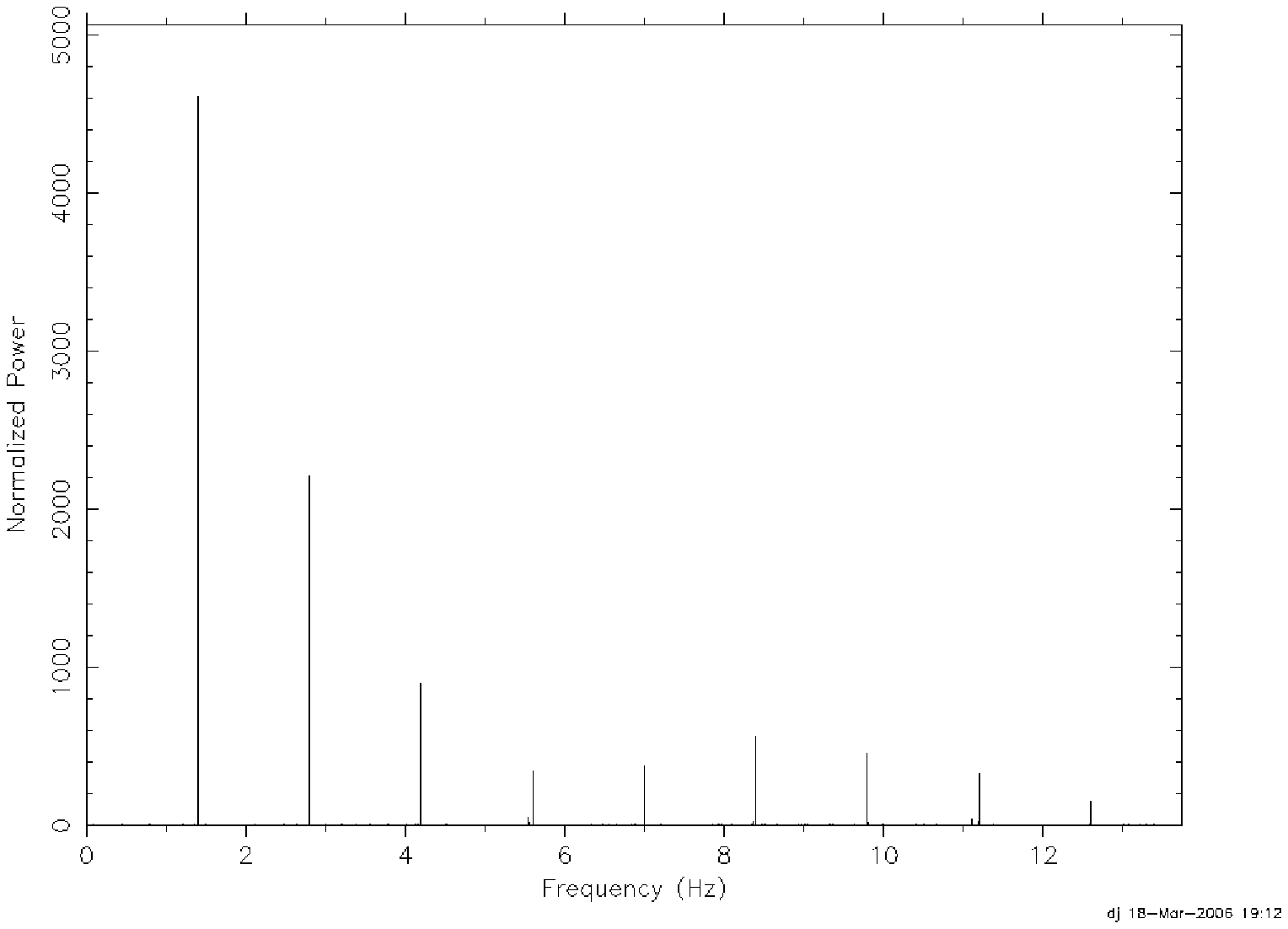}
\\
\caption{FFT spectra of PSR B0329+54 after summing two polarisations off line.}
\label{fft}
\end{figure}
\end{center}

\section{Primary Results and Prospect}
\subsection{Primary Results}
We have set up a data acquisition system at Urumqi Observatory. Observations for know pulsars show that after summing two polarisations off line, our data can perform pulsar searching use Sigproc \&  Presto. Fig. 1 and Fig. 2 present the de-dispersion and Fourier transform results on PSR B0329+54 using Presto software.  

\subsection{Prospect}
Some 3EG sources (Hartman et al. 1999) and one Tev source were observed, the data reduction are continuing at present. It needs ten points to cover  $\sim 1.5^{\circ}$ error box of for each source for the beam size of $0.5^{\circ}$ at 18~cm (Lorimer \& Kramer, 2005). Every pointing lasts at least 5-hour  observation. Recording format is similar to Parkes’. 

The 3EG catalog was inspected to look for candidate sources satisfying the following criteria:

(1) They must not be listed as identified sources in the catalog.

(2) They should be located at galactic latitudes within the following band: $ 5^{\circ} < |b| < 30^{\circ} $ .

(3) They should have hard spectra, with $\gamma$-ray photon indices  $ \Gamma < 2 $, within the errors quoted in the catalog.

(4) They should be non-variable sources according to the main variability indices introduced in the literature: the I index (Torres et al.,2001a) and the $\tau $ index (Tompkins 1999). These two indices are in general well-correlated, at a  $ 7- \sigma $ level (Torres, et al.,2001b). 

(5) They do not exist in the lists of the other group (Champion et al.,2005, Kramer et al., 2003, Roberts et al.,2004, Camilo et al.,2001, D'Amico et al.,2001, Torres et al.,2001, Halpern et al.,2001, Roberts et al.,2002, Becker et al.,2004).

Now big radio telescopes including ALFA (Cordes et al., 2006), GBT (Ransom, 2005), updated Lovell Telescope in the northern Hemisphere are active for pulsar survey. With our telescope, the advantage of adequate telescope time might give us a chance to find pulsars in target searching. It is also possible to find transient radio sources (Cordes, J.M., et al., 2004, Mclaughlin M.A., 2001, O'Brien,J. T., et al., 2005), especially those having relatively high flux density (Hyman, S.D., et al 2005). Additionally, GLAST will find hundreds of Gamma-ray sources(McLaughlin \& Cordes, 2000) which indicates an increasing searing cadidates.

It will be excellent to find a pulsar using domestic radio telescope. However gaining experience from this system is an important goal. Big radio telescopes, such as FAST, 50~m of Miyun telescope etc will have better chance in finding pulsars. 

\begin{acknowledgements}
We thank the engineers responsible for maintaining the receiver, telescope at Urumqi Observatory, and the staffs who helped with the observations. DJ thanks humourous Dr Dunc Lorimer and sharply Dr Scott M Ransom give me kindly help about software and advice about our search system. This program is supported by Key Directional Project ***  of CAS.
\end{acknowledgements}

\label{lastpage}
\end{document}